\documentclass[twocolumn]{aastex631}

\newcommand{\noun}[1]{\textsc{#1}}

\begin{document}
\title{X-ray Polarimetry as a Tool to Constrain Orbital Parameters in X-Ray Binaries}

\author[0000-0002-9774-0560]{John Rankin}
\affiliation{INAF Istituto di Astrofisica e Planetologia Spaziali, Via del Fosso del Cavaliere 100, 00133 Roma, Italy}

\author[0000-0002-7502-3173]{Vadim Kravtsov}
\affiliation{Department of Physics and Astronomy, FI-20014 University of Turku, Finland}

\author[0000-0003-3331-3794]{Fabio Muleri}
\affiliation{INAF Istituto di Astrofisica e Planetologia Spaziali, Via del Fosso del Cavaliere 100, 00133 Roma, Italy}

\author[0000-0002-0983-0049]{Juri Poutanen}
\affiliation{Department of Physics and Astronomy, FI-20014 University of Turku, Finland}

\author[0000-0003-4952-0835]{Fr\'{e}d\'{e}ric Marin}
\affiliation{Universit\'{e} de Strasbourg, CNRS, Observatoire Astronomique de Strasbourg, UMR 7550, 67000 Strasbourg, France}

\author[0000-0002-6384-3027]{Fiamma Capitanio}
\affiliation{INAF Istituto di Astrofisica e Planetologia Spaziali, Via del Fosso del Cavaliere 100, 00133 Roma, Italy}

\author[0000-0002-2152-0916]{Giorgio Matt}
\affiliation{Dipartimento di Matematica e Fisica, Universit\`a degli Studi Roma Tre, Via della Vasca Navale 84, 00146 Roma, Italy}

\author[0000-0003-4925-8523]{Enrico Costa}
\affiliation{INAF Istituto di Astrofisica e Planetologia Spaziali, Via del Fosso del Cavaliere 100, 00133 Roma, Italy}

\author[0000-0003-0331-3259]{Alessandro {Di Marco}}
\affiliation{INAF Istituto di Astrofisica e Planetologia Spaziali, Via del Fosso del Cavaliere 100, 00133 Roma, Italy}

\author[0000-0003-1533-0283]{Sergio Fabiani}
\affiliation{INAF Istituto di Astrofisica e Planetologia Spaziali, Via del Fosso del Cavaliere 100, 00133 Roma, Italy}

\author[0000-0001-8916-4156]{Fabio {La Monaca}}
\affiliation{INAF Istituto di Astrofisica e Planetologia Spaziali, Via del Fosso del Cavaliere 100, 00133 Roma, Italy}
\affiliation{Dipartimento di Fisica, Università degli Studi di Roma “Tor Vergata”, Via della Ricerca Scientifica 1, 00133 Roma, Italy}
\affiliation{Dipartimento di Fisica, Università degli Studi di Roma “La Sapienza”, Piazzale Aldo Moro 5, 00185 Roma, Italy}

\author[0009-0001-4644-194X]{Lorenzo Marra}
\affiliation{Dipartimento di Matematica e Fisica, Universit\`a degli Studi Roma Tre, Via della Vasca Navale 84, 00146 Roma, Italy}

\author[0000-0001-8916-4156]{Paolo Soffitta}
\affiliation{INAF Istituto di Astrofisica e Planetologia Spaziali, Via del Fosso del Cavaliere 100, 00133 Roma, Italy}

\begin{abstract}
X-ray binary systems consist of a companion star and a compact object in close orbit. Thanks to their copious X-ray emission, these objects have been studied in detail using X-ray spectroscopy and timing.
The inclination of these systems is a major uncertainty in the determination of the mass of the compact object using optical spectroscopic methods. In this paper, we present a new method to constrain the inclination of X-ray binaries, which is based on the modeling of the polarization of X-rays photons produced by a compact source and scattered off the companion star. We describe our method and explore the potential of this technique in the specific case of the low mass X-ray binary GS~1826$-$238 observed by the Imaging X-ray Polarimetry Explorer (IXPE) observatory.
\end{abstract}

\section{Introduction}

X-ray binary systems are among the brightest celestial objects in the X-rays. They are powered by mass transfer from a companion star to a compact object, either a white dwarf, neutron star or a black hole, which orbits at a short distance. Their high luminosity allowed the discovery and the detailed study of these systems since the dawn of X-ray astronomy. To achieve this, spectroscopy and timing were critical tools. 

One of the most interesting parameters of these systems is the mass of the compact object, which, for example, can constrain the equation of state of ultra–dense matter in neutron stars \citep{Miller2020} or clarify the origin of intermediate mass black holes which are observed as sources of gravitational waves \citep{Mehta2022}. However, apart from very special systems like double pulsars, measuring the mass in X-ray binaries is challenging because these systems currently cannot be spatially resolved. In some cases, the radial velocity can be measured both for the companion star through optical observations and for the compact object with, e.g., X-ray pulsations. Even in those conditions, one can only derive the ratio of the mass of two objects; determining the individual masses requires solving the mass function, which depends on the orbital inclination of the system.

The inclination of the orbit can be constrained when the source shows eclipses or dips, which can be due to obscuration of the central X-ray source by the companion; however, this requires the system to be nearly edge-on. Another possibility is in systems that accrete via Roche lobe, where the shape of the companion star can be significantly distorted due to the Roche lobe geometry; in this case, the size of the surface visible to the observer changes with the orbital phase in an inclination-dependent way, creating a modulation in the optical-light emission \citep{1997ApJ...477..876O}.

In addition to the importance of fixing the parameters of the binary system, the orbital inclination is also of interest per se. For example, it can be compared with the orientation of the accretion disk if the latter is known from other means, e.g., optical polarimetry.
This allows to verify if these two are aligned or not, with important constraints on the geometry of the mass transfer. 

In this paper, we investigate the possibility of determining the orbital inclination, and potentially other orbital parameters, by measuring the polarization of the X-ray photons generated in the vicinity of the compact object (e.g., from the accretion disk, corona and/or spreading/boundary layer in the case of a neutron star) and then scattered off the companion star outer envelopes. It has been speculated a long time ago that the effect should be detectable as a polarization $\sim$1\% \citep{Basko1974,Rees1975}, but measuring polarization in the X-rays is notoriously challenging. In fact, before the launch of the NASA/ASI Imaging X-ray Polarimetry Explorer (IXPE) mission \citep{Weisskopf2022, 2021AJ....162..208S}, only very few detections of X-ray polarization had been made.

The approach discussed here has been attempted at other wavelengths, e.g., at optical wavelengths \citep{2020A&A...643A.170K}. Optical polarimetric observations, however, have some uncertainties due to the high contamination from interstellar dust polarization.
Furthermore, the emission in the vicinity of the compact source is much lower at optical wavelengths than in the X-rays and, in low-mass X-ray binaries (where the mass of the companion is $<$~1 solar mass), the orbital variability is lost in the variability of the much brighter accretion disk.

In this paper we look for this polarization signal in the IXPE observation of GS~1826$-$238, which is a low-mass X-ray binary hosting a weakly magnetized neutron star. This is an ideal candidate for our study:  \citet{Capitanio2023} studied the polarization of the source itself --- due to the accretion geometry of the disc/corona --- finding it to be unpolarized at high significance.  As a consequence any phase resolved polarization we observe is due to orbital effects --- and the short orbital period ($\sim$2.2~hr, \citealt{2010AstL...36..738M}) allows us to average intrinsic source variations over several orbits, even during relatively short observations. 

This paper is structured as follows. We first describe the orbital polarization model in Section \ref{sec:model}, and then present the expected polarization as a function of the orbital phase in Section~\ref{sec:variation_with_phase}. 
In Section~\ref{sec:observation} we apply this to the GS~1826$-$238 system. Finally, in Section~\ref{sec:conclusion}, we present our conclusions.

\section{Modelling Polarization Induced by Orbital Scattering\label{sec:model}}

\begin{figure}
\centering
\includegraphics[width=0.9\linewidth]{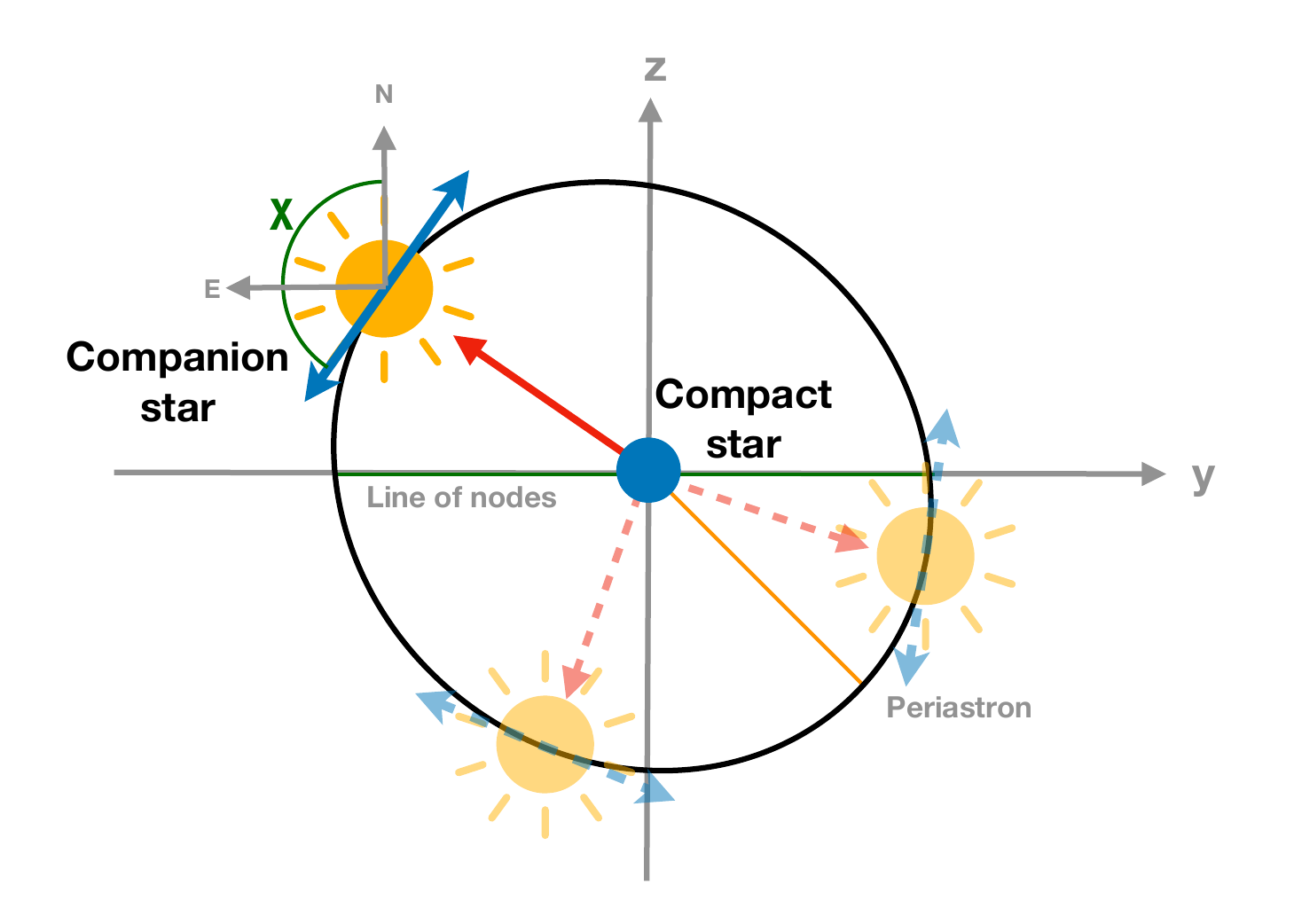}
\includegraphics[width=0.9\linewidth]{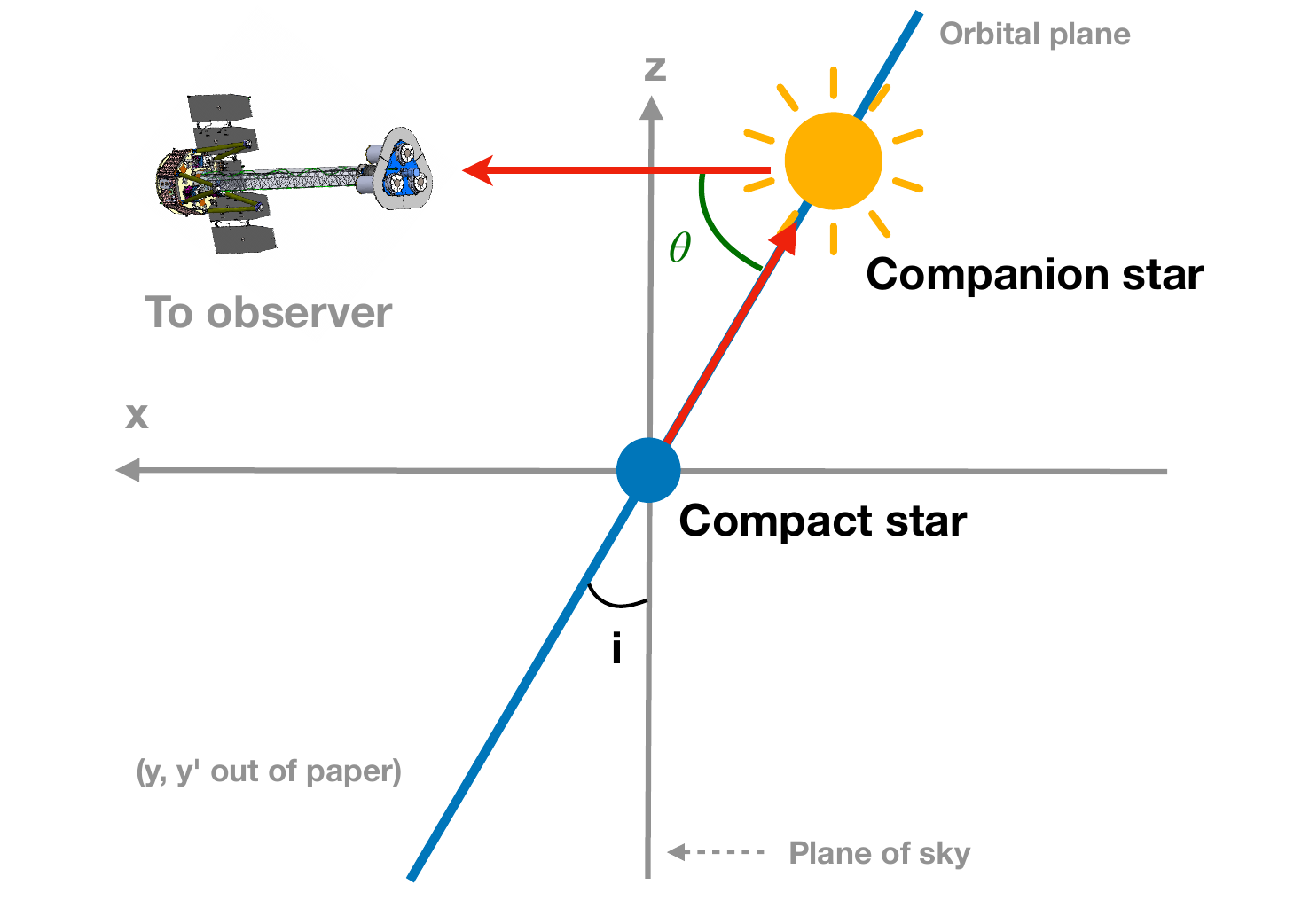}
\caption{Representation of the geometry of the orbital model (in a reference system centered on the compact object such that the $z$ axis points to the celestial north, the $y$ axis west, and the $x$ axis towards the observer). (Top) View seen from the observer's point of view. The central compact object is the blue central circle, while the orange companion star orbits it. The radiation emitted from the compact object is reflected by the companion star, and is this way polarized. 
In Compton scattering the direction of the scattered photons is orthogonal to the scattering plane; the blue lines identify such a direction for a distant observer, corresponding to the direction of polarization.
The polarization angle is given by $\chi$. (Bottom) Reference system used to represent orbits. The polarization degree depends on the scattering angle $\theta$.}
\label{fig:model_geometry}
\end{figure}

To quantitatively investigate the information which can be obtained by X-ray polarization, we developed a simple model which predicts the polarization as a function of the orbital parameters. 
We modelled the orbital motion of the compact object and companion star with the keplerian \noun{TwoStars} code that comes with \citet{2006ima..book.....C}, which we ported into \noun{python}. This code computes the orbits using the Kepler equation, in which the orbital distance as a function of the orbital plane longitude $\lambda$ is given by
\begin{equation}
    r(\lambda) = \frac{a(1-e^2)}{1+e\cos(\lambda - \lambda_p)}
\end{equation}
where $a$ is the semi-major axis of the orbit and $\lambda_p$ the longitude of periastron. The orbit is described in a reference system such that the $z$ axis points to the celestial north, the $y$ axis west, and the $x$ axis towards the observer, the orbital angular momentum is inclined to the line-of-sight at an angle $i$.
The reference system is represented from the side in Figure~\ref{fig:model_geometry} (bottom), and from the observer's point of view in Figure~\ref{fig:model_geometry} (top).

\begin{figure*}
\centering
\includegraphics[width=0.8\linewidth]{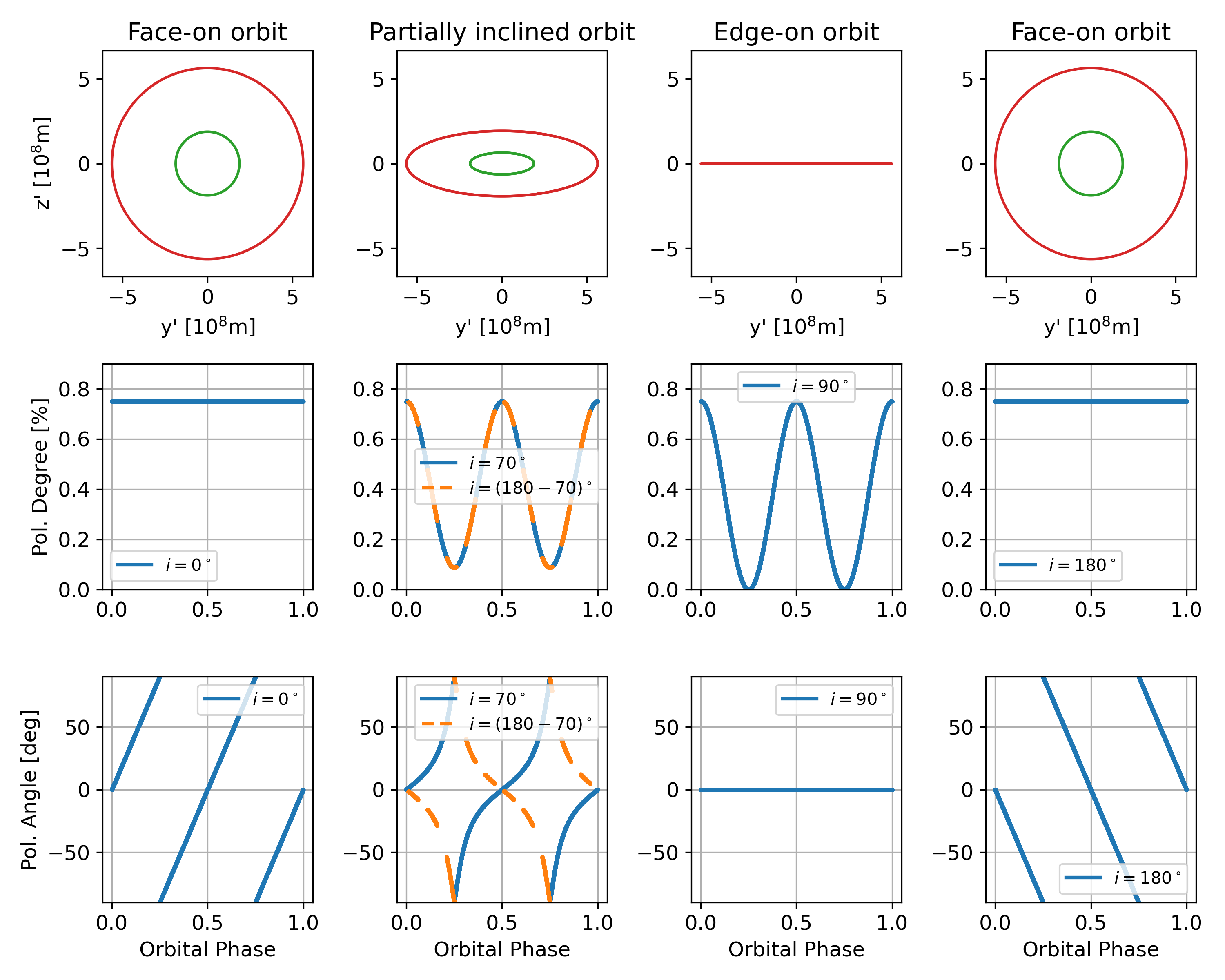}
\caption{Polarization derived from the orbital model as a function of the orbital phase. The top row shows the projection of the orbital configuration seen from the observer's point of view, while two lower rows show the polarization degree and angle. In these examples, to avoid discontinuities in the polarization angle, the argument of periapsis is set to $\omega=0\degr$ and the longitude of the ascending node is set to $\Omega=90\degr$; the scattering fraction is set to $f_{\rm sc}=0.02$ and the eccentricity is set to $e=0$ for graphical simplicity.
(Left) Face-on orbit. In this case $\theta$ is constant and so the polarization degree is also constant, while the polarization angle depends on the orbital phase. (Center left) Inclinations of  $70\degr$ 
and $(180-70)\degr$.  The first case is counterclockwise and the second is clockwise.
(Center right) Edge-on orbit. In this case the polarization degree depends on the orbital phase, while the polarization angle is constant. (Right) Face-on orbit but counterclockwise with respect to that on the left column ($i=180\degr$)}
\label{fig:orbit-examples}
\end{figure*}

The polarization due to scattering is computed as in appendix A of \cite{2020A&A...643A.170K}, where elastic Thomson scattering of photons is assumed to occur on the companion star. The shape of the observed normalized Stokes $q$ and $u$ profiles is determined by the geometry of the orbit (inclination, eccentricity $e$ and the orientation of the orbit on the sky), while the amplitude depends on the fraction of scattered radiation $f_{\rm sc}$, which in turn depends on the total number of electrons in the cloud $N_{\rm e}$ and the binary separation $r$ as $f_{\rm sc}=N_{\rm e}\sigma_{\rm T}/(4\pi r^2)$, where $\sigma_{\rm T}$ is the Thomson cross-section. Electron temperature does not affect the polarization because it is relatively low in the stellar atmosphere so that the scattering can be considered in the Thomson regime. In this approach the polarization is given by
\begin{equation}
P = \frac{3}{8} f_{\rm sc} \sin^2 \theta , 
\end{equation}
where $\theta$ is the scattering angle (shown in Figure~\ref{fig:model_geometry}). 

\section{Polarization as a Function of Orbital Phase \label{sec:variation_with_phase}}

\subsection{Polarization for Different Orbital Inclinations}

We show four scenarios in Figure~\ref{fig:orbit-examples} representing a face-on, an inclined, an edge-on and a clockwise face-on orbit. In all cases the polarization degree is lower than 1\%, but the variation of either the polarization degree, angle or both depends on the orbit parameters. In particular:
\begin{itemize}
\item Face-on orbit (left column of the figure): the polarization angle varies with the orbital phase, while the polarization degree is constant because the scattering angle $\theta$ is constant.
\item Intermediate inclinations (in the center of the figure): for these inclinations  the minimum of the polarization degree is higher than in edge-on orbits, but the maximum is the same.  The direction of the orbital motion, counterclockwise (i.e. $i<90\degr$) or clockwise (i.e. $i>90\degr$) as seen from Earth,  causes opposite  trends in the polarization angle. 
\item Edge-on orbit (right column of the figure): the polarization degree varies between 0 and  its maximum depending on the orbital phase, while the polarization angle is constant.
\end{itemize}

\begin{figure*} 
\centering
\includegraphics[width=0.8\linewidth]{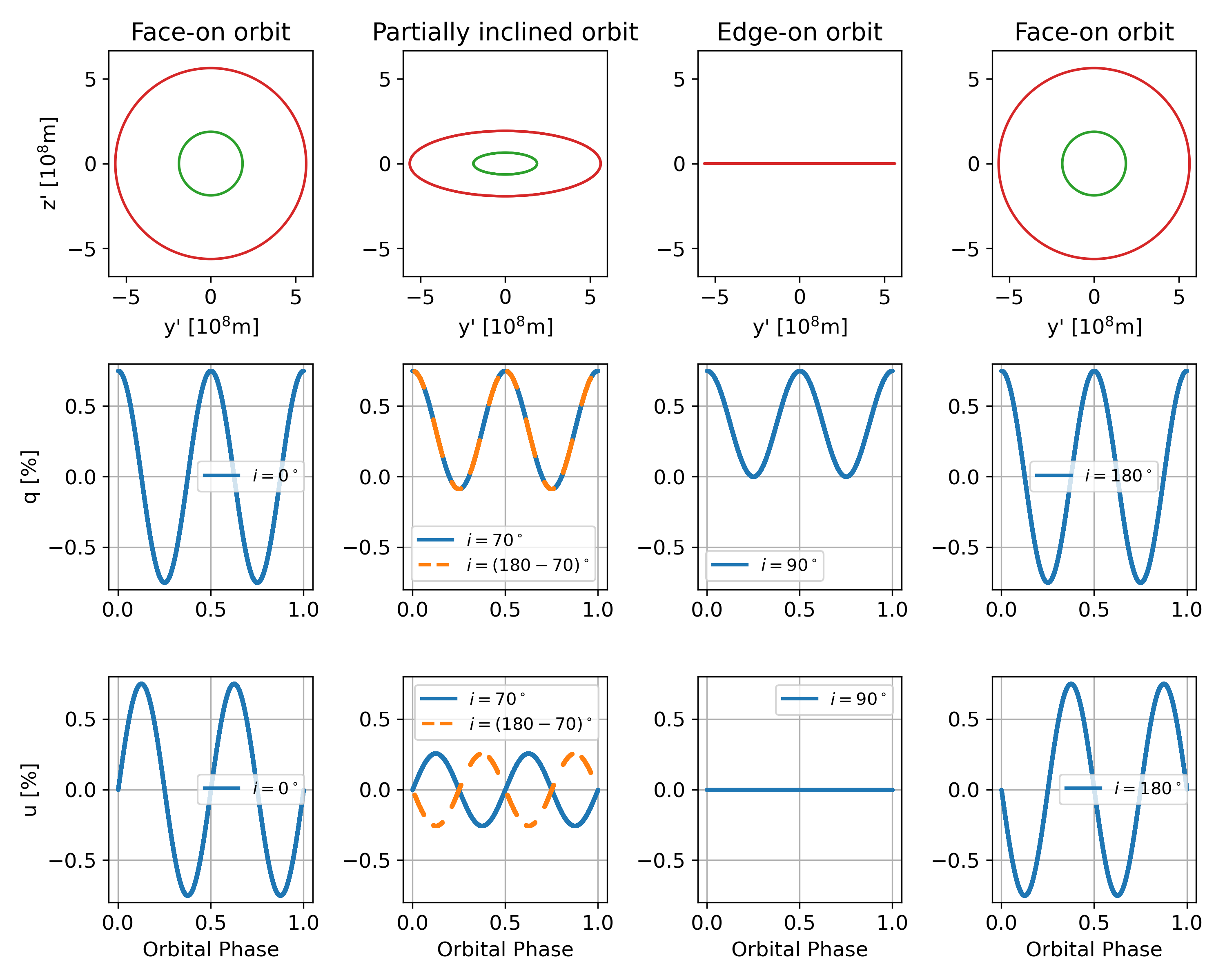}
\caption{Same as Figure~\ref{fig:orbit-examples} but for the normalized Stokes parameters.}
\label{fig:orbit-examples-Stokes}
\end{figure*}

Linear polarization is often represented not as polarization degree and angle, but through the normalized Stokes parameters $q$ and $u$  \citep{Trippe2014, Kislat2015}, which have the advantage of being statistically independent. Because these are the parameters used in the rest of this paper, Figure~\ref{fig:orbit-examples-Stokes} represents the same configurations as Figure~\ref{fig:orbit-examples} but with the Stokes parameters.
We see that a change in the sense of rotation 
causes an inversion in $u$ but not in $q$.

\subsection{Variations of Other Orbital Parameters}

\begin{figure}
\centering
\includegraphics[width=0.9\linewidth]{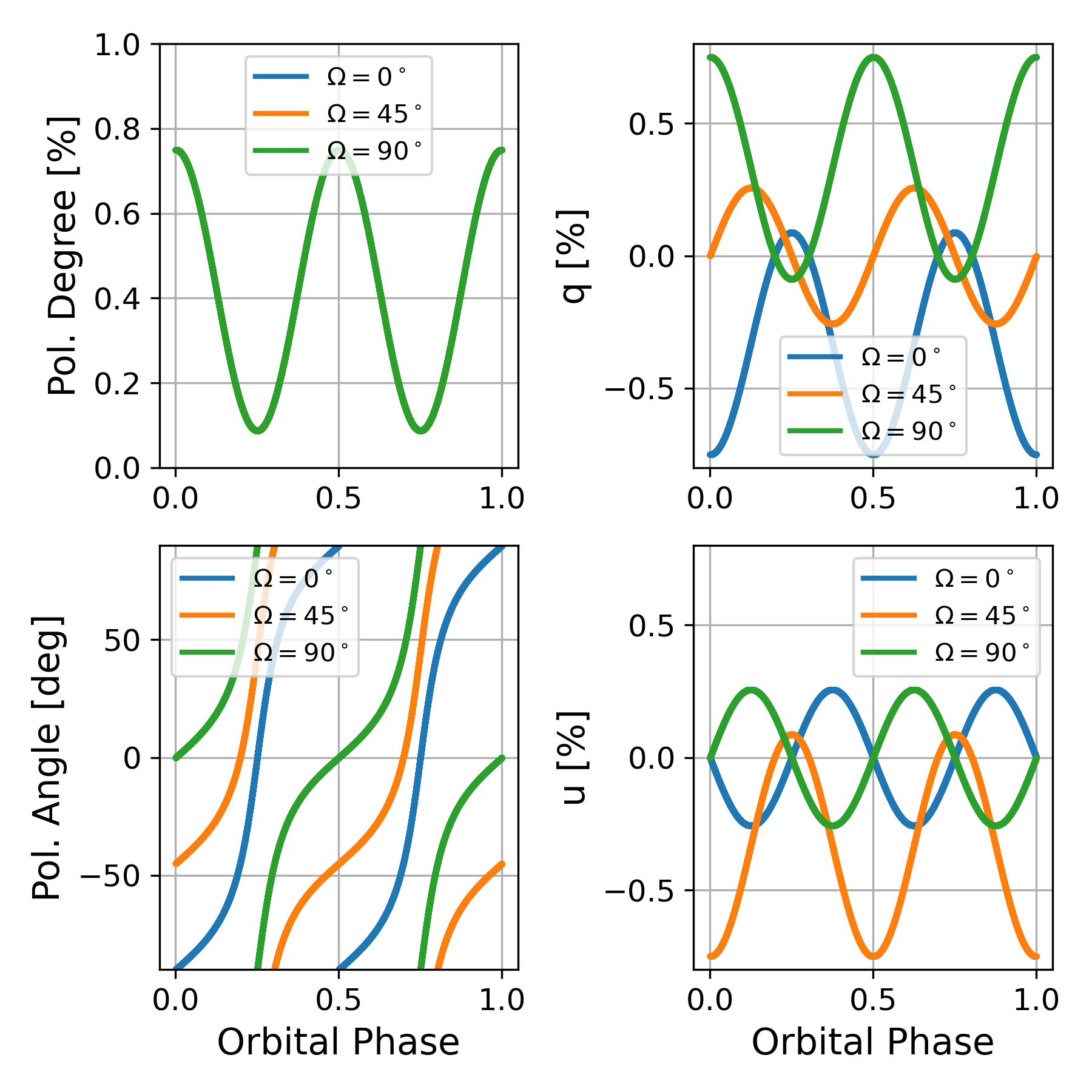}
\caption{Polarization degree, angle and Stokes parameters, as a function of orbital phase, for different longitudes of ascending node $\Omega$. The other parameters are set to $i=70\degr$ and $\omega=0\degr$. The change of $\Omega$ is in fact a standard rotation of the Stokes parameters by the rotation matrix with the argument equal to $2\Omega$.
} 
\label{fig:long-node-phase-trend}
\end{figure}

Because low mass X-ray binaries are very old, the orbit has stabilized as a circular orbit \citep{Lecar1976}, so that we can set the eccentricity to $e=0$; the remaining parameters that influence the trend of polarization with orbital phase are the argument of periapsis $\omega$ and the longitude of ascending node $\Omega$.
Because $e=0$, $\omega$ produces only a phase shift in the trend of polarization with orbital phase. 
Figure~\ref{fig:long-node-phase-trend} shows instead the polarization trend with orbital phase for different $\Omega$: the polarization degree is not sensitive to $\Omega$, while the effect on $q$ and $u$ is more evident --- varying $\Omega$ causes the variations with phase to be more visible in one parameter or in the other.

\section{Searching for Orbital Polarization in the GS~1826$-$238 System \label{sec:observation}}

\subsection{IXPE observation of GS~1826$-$238}
The IXPE observatory \citep{Weisskopf2022} consists of three identical X-ray mirror modules, with at their focus three identical X-ray polarization sensitive Gas Pixel Detectors \citep{2021AJ....162..208S,2021APh...13302628B}. These detectors measure the energy, position, time and linear polarization of the incident X-rays, in the 2--8~keV energy range.

The IXPE observation of GS~1826$-$238 lasted for $\sim$90~ks and was taken on 2022 March 29--31. The orbital-averaged results have already been analyzed by \citet{Capitanio2023}; here we focus on the analysis of the variation of polarization along the orbit. 

Data files, reduced by the standard IXPE pipeline running at Science Operations Center in NASA/MSFC, were downloaded from the IXPE public archive at HEASARC.\footnote{https://heasarc.gsfc.nasa.gov/docs/ixpe/archive/} (obs id 01002801 v.03) Reduced data are corrected for temporal gain variations, which are monitored during the observation with in-flight calibration sources \citep{Ferrazzoli2020}, and for the response to unpolarized radiation \citep{2022AJ....163...39R}. Event-by-event Stokes parameters are calculated following \cite{Kislat2015} and computed using the weighted scheme described by \citet{2022AJ....163..170D}; they are provided to the user in a reference frame projected on the sky. 

We analyzed these data files using \textsc{ixpeobssim} 30.6.4 \citep{Baldini2022},  \textsc{heasoft} 6.32.1 \citep{2014ascl.soft08004N} and \textsc{xspec} 12.13.1 \citep{Arnaud1996}.
To exclude the background, we selected a circular region of radius 115\arcsec\ around the source using the \textsc{SAOimageDS9} software \citep{2003ASPC..295..489J}. We used the \texttt{barycorr} \texttt{FTOOL} in \textsc{heasoft} to convert photon arrival times to the Solar System Barycenter.

\begin{figure}
\centering
\includegraphics[width=1\linewidth]{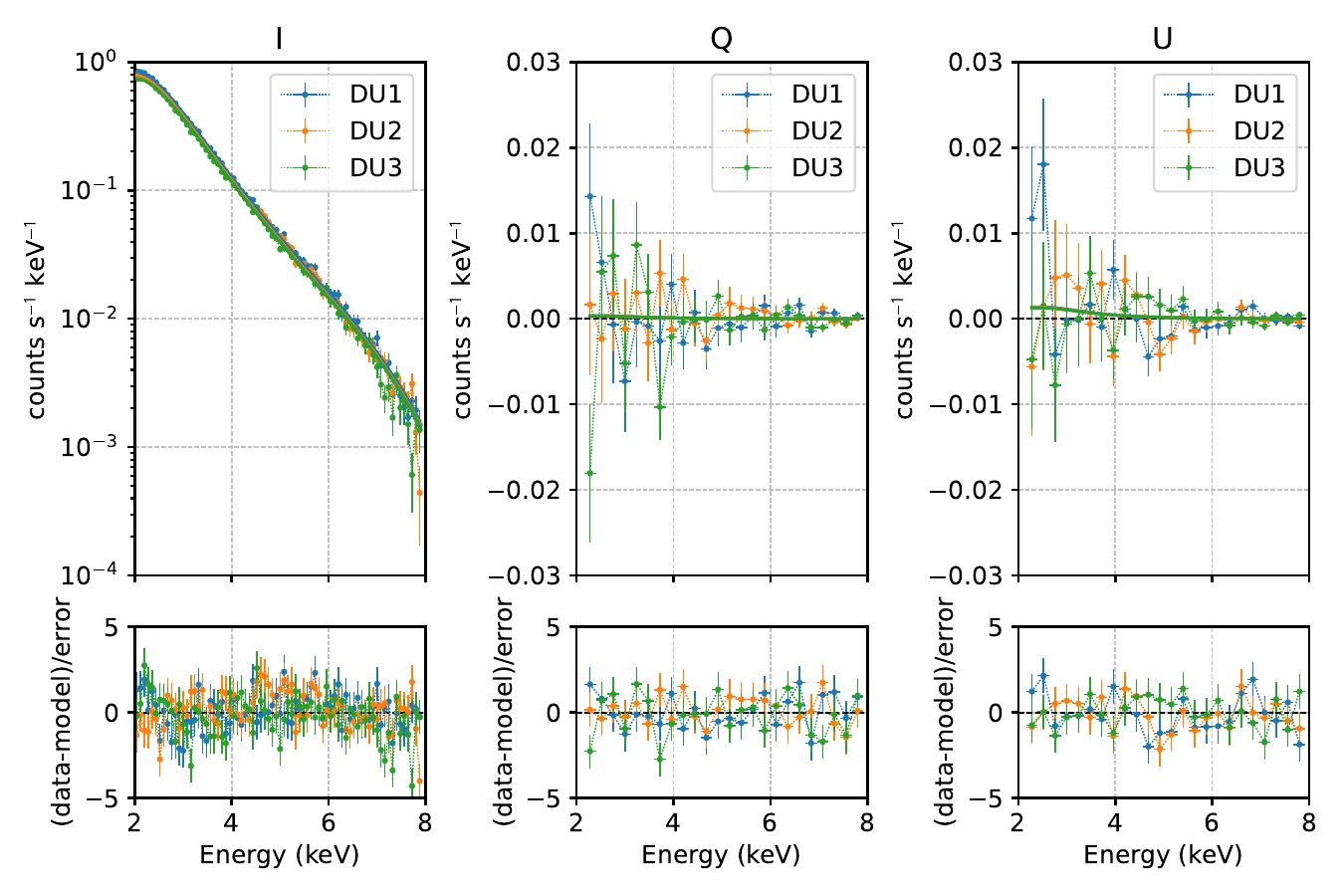}
\caption{Example \textsc{xspec} fits of the $I$, $Q$, and $U$ Stokes parameters in the phase bin 5 (the others are similar). 
Blue, red, and green crosses correspond to the IXPE detector units (DUs) 1, 2, and 3, respectively.
As also reported in  \citet{Capitanio2023}, the observation was carried out with the mirrors slightly offset with respect to the nominal position. This caused some uncertainties in the response functions which are yet to be modelled, and contributes to the small residuals present in the fit, without affecting the measured polarization \citep{Capitanio2023}.
} 
\label{fig:xspec_example}
\end{figure}

We divided the observation in different orbital phase bins, each event being assigned to the appropriate bin by the \texttt{xpphase} tool of \textsc{ixpeobssim}.
We set the folding frequency to that corresponding to the orbital period of 2.2494~h (found by \citealt{2010AstL...36..738M}). The derivatives of the folding frequency were set to 0. We then obtained the binned spectra for Stokes parameters $I$, $Q$, and $U$ using the \texttt{xpbin} tool of \textsc{ixpeobssim}.

\begin{figure}
\begin{centering}
\includegraphics[width=0.9\linewidth]{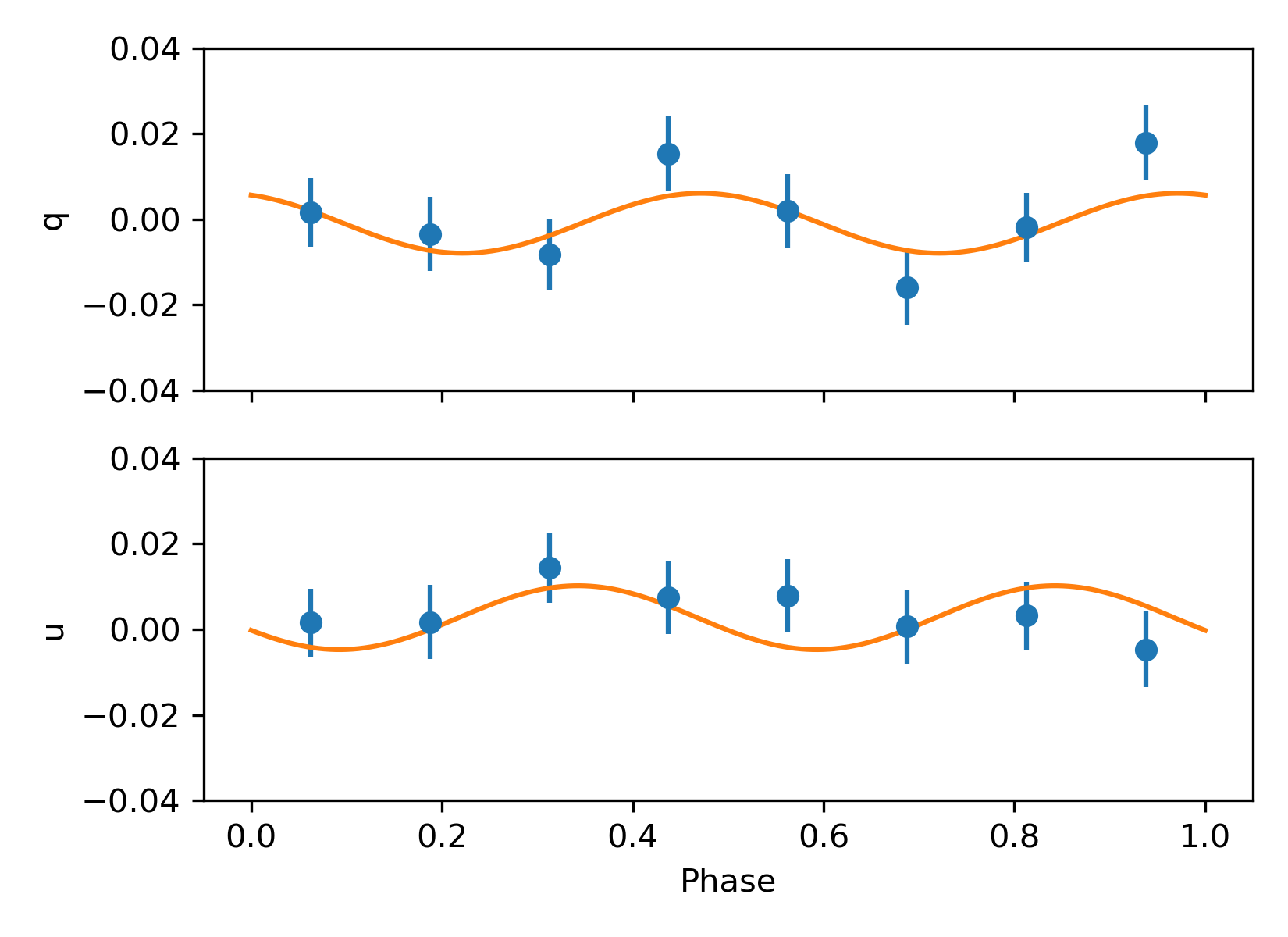}
\end{centering}
\caption{Orbital dependence of the normalized Stokes parameters $q$ and $u$ for  GS~1826$-$238 and the best-fit model described in Section \ref{sec:model}. 
The error bars are 1$\sigma$. 
The best-fit has $\chi^2=8.7$ for 12 d.o.f., while a  fit with a constant gives $\chi^2=15.9$ for 14 d.o.f.
Akaike Information Criterion (AIC) test gives a value of $AIC=2261$ for the orbital model and $AIC=5392$ for the constant fit, indicating that the fit with our model is more significant than with a constant.
\label{fig:Fit-results}}
\end{figure}

We obtained the normalized Stokes parameters $q=Q/I$ and $u=U/I$  using \textsc{xspec}.
We defined the user-defined polarization model  \texttt{stokesconst}, in which the $q$ and $u$ Stokes parameters are fit as constants as a function of energy; compared to \texttt{polconst} this has the advantage of not having the shift in polarization angle between $90\degr$ and $-90\degr$. We fit the data in the 2--8~keV band using the spectral model used by \citet{Capitanio2023} multiplied by \texttt{stokesconst}: \texttt{tbabs*(diskbb+comptt)*stokesconst}, the only parameters left free to vary were $q$ and $u$. 
Figure~\ref{fig:xspec_example} shows an example spectral fit.
The dependencies of  Stokes parameters $q$ and $u$ on orbital phase are shown in Figure~ \ref{fig:Fit-results}. 

\begin{figure*}
\centering
\includegraphics[width=0.8\linewidth]{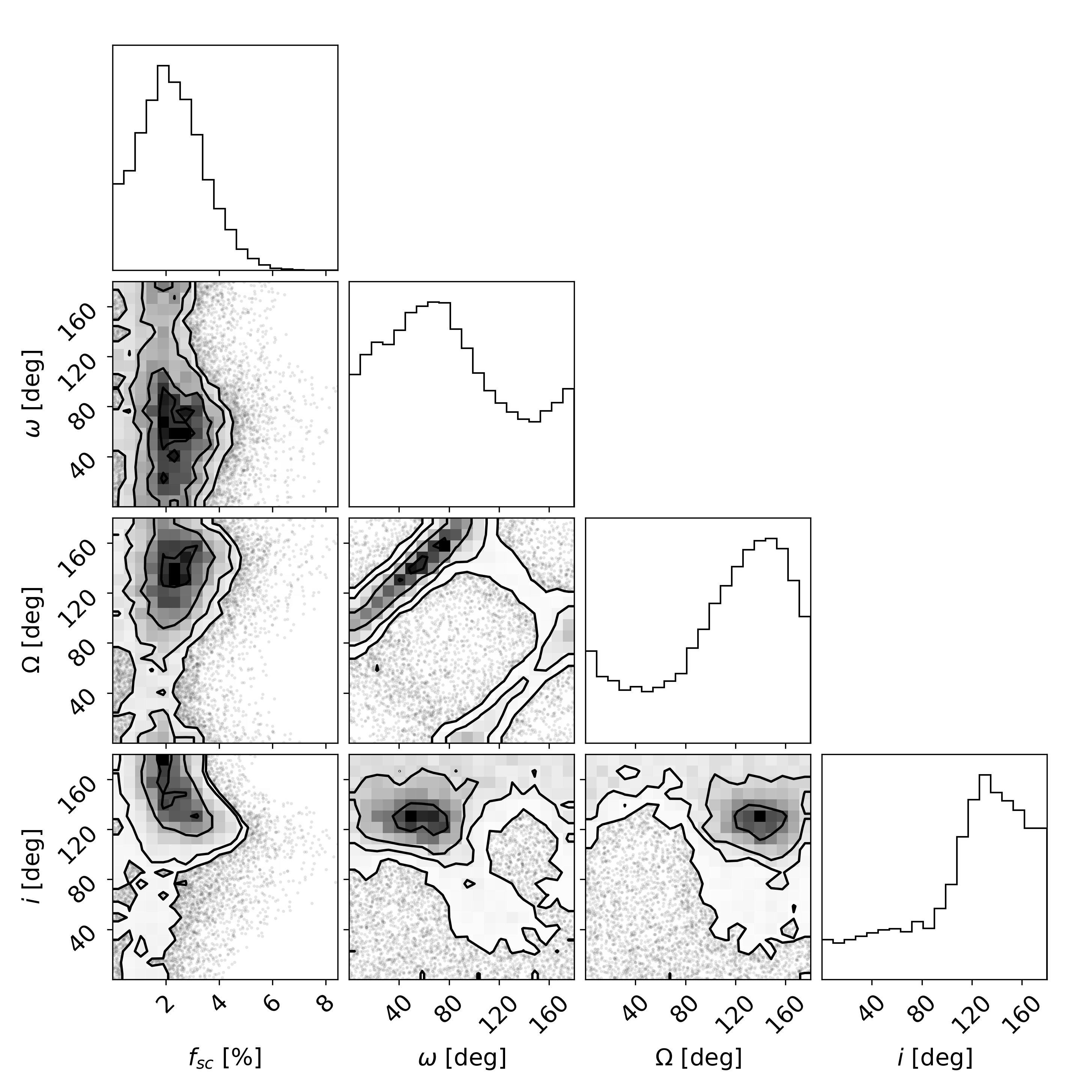}
\caption{
Posterior probability distribution of the reflection model parameters: the scattering fraction $f_{\rm sc}$, the argument of periapsis $\omega$, the longitude of ascending node $\Omega$, and the inclination $i$ from the fitted model; $\omega$ and $\Omega$ are partially degenerate because the orbit is not edge-on. The contour levels are 0.5, 1, 1.5, 2 $\sigma$.
\label{fig:corner}}
\end{figure*}

Compared to the analysis by \citet{Capitanio2023}, our analysis uses the same data divided into 8 phase bins: as a consequence we expect a $\sqrt{8}$ reduction in sensitivity in each bin; the actual reduction is 10\%--15\% smaller because the analysis in this paper uses the weighting scheme by \citet{2022AJ....163..170D}, while \citet{Capitanio2023} do not. Because the polarization angle rotates in the different time bins, when doing a phase-averaged analysis (as done by \citealt{Capitanio2023}) the polarization of the different bins cancels out. The analysis carried out by \citet{Capitanio2023} investigated the structure of the corona (close to the neutron star), while our analysis, being resolved in phase with the orbital period, is sensitive to orbital phenomena on a much larger scale.

\subsection{Fitting Polarization as a Function of the Orbital Phase}

We fit the folded data to the orbital model described in Section \ref{sec:model}; the $q$ and $u$ dependencies were fitted simultaneously, keeping as free parameters the inclination $i$, $\omega$, and $\Omega$. ~
We minimized the $\chi^2$ of the fit and derived the posterior probability distributions for the parameters of the model with the nested sampling Monte Carlo algorithm
\noun{MLFriends} \citep{2016S&C....26..383B, 2019PASP..131j8005B} using the
\noun{UltraNest}\footnote{\url{https://johannesbuchner.github.io/UltraNest/}} package \citep{2021JOSS....6.3001B}. 
Figure~\ref{fig:corner} shows the obtained distributions and  the best-fit model is shown in Figure~\ref{fig:Fit-results}. 
The numerical values of the fitted parameters and their uncertainties are reported in Table~\ref{tab:results}.
The orbital inclination is $132 ^{+47}_{-24}$~deg, which is inside the interval from $90\degr$ to $180\degr$, indicating that the stars rotate in the clockwise direction on the sky. 

\begin{deluxetable}{ccc}
\tablecaption{Best-fit parameters for the reflection model. \label{tab:results}}
\tablehead{
\colhead{Parameter} & \colhead{Units} & \colhead{Value} 
}
\startdata  
$f_{\rm sc}$ & (\%) & $2.7^{+1.0}_{-1.2}$ \\
$i$ & (deg) & $132^{+47}_{-24}$\\
$\omega$ & (deg) & $57^{+42}_{-45}$\\ 
$\Omega$ & (deg)  & $144^{+34}_{-45}$ \\
\enddata
\tablecomments{Uncertainties are at 68\% confidence level (1$\sigma$).}
\end{deluxetable}

\begin{figure}
\centering
\includegraphics[width=0.75\linewidth]{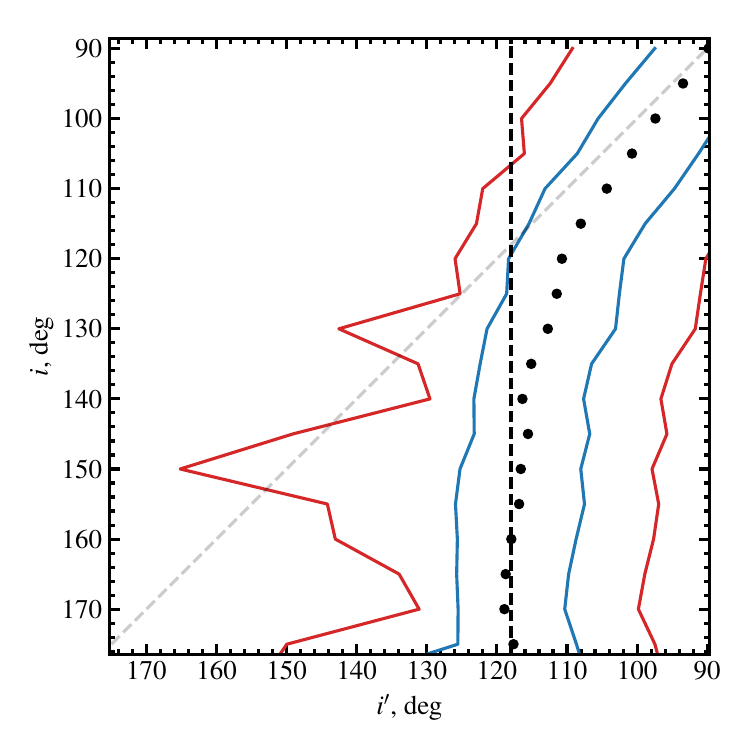}
\caption{Relation between the real inclination ($i$) and the inclination obtained from the fit ($i'$), shown with black circles. 
The blue and red lines are the 1$\sigma$ and 2$\sigma$ confidence intervals.
The dashed vertical line shows the critical inclination of about 120\degr.
\label{fig:inclination-confidence}}
\end{figure}

The uncertainties obtained above (Table~\ref{tab:results}) do not correspond to the actual confidence intervals; in particular the inclination is biased towards higher values \citep{1994MNRAS.267....5W}. To estimate the correct intervals, following the same procedure as \cite{Kravtsov2023}, we modelled orbital variations of the Stokes parameters for inclinations in the range from 90$\degr$ to 180$\degr$, with a step of 5$\degr$, for the same phases of the observations as in the data. Next, we added Gaussian noise to these points with the same $\sigma$ as the data and, after that, we fit the model back to the simulated data. The result is a relation between the real inclination $i$ and that estimated from the formal fit ($i'$), shown in Figure~\ref{fig:inclination-confidence}.
From the simulations we get a critical inclination of about 120$\degr$, implying that we are only sensitive to inclinations in the range from 90$\degr$ to 110$\degr$, while any fit of an inclination value greater than the critical inclination will result in an inclination of about 120$\degr$. The value of inclination we obtain above from the fit is close to this critical inclination, indicating that we have weak constraints on the inclination: its value could likely be higher than found from the confidence interval of Table \ref{tab:results}, covering the entire interval up to 180$\degr$.

The large estimated range of the orbital inclination with respect to the observer, if one allows for the indetermination of the direction of motion, is consistent with the previous estimates of \citet{Johnston2020}, who found an inclination of $69^{+2}_{-3}$~deg by modeling multi-epoch X-ray bursts, and of \citet{2011AstL...37..826M}, who found an inclination of $62\fdg5\pm5\fdg5$ by looking at modulations in the optical flux.
 
\section{Conclusion}
\label{sec:conclusion}

We presented a method to study the orbits of X-ray binaries using X-ray polarization, which is based on the study of the X-ray photons scattered off the companion star. A basic model is built to estimating the polarization as a function of the orbital phase: this trend depends on the orbital inclination and on other orbital parameters.

We searched for this orbital polarization trend in GS~1826$-$238, a low mass X-ray binary system which was observed by IXPE with a short pointing in 2022. The statistical uncertainty on the constraints we find when fitting the model is very large, but the results are not incompatible with an inclination exceeding 90\degr.  

Future longer observations of GS~1826$-$238 could bring better statistics to constrain the orbital parameters.
Observations of other X-ray binary systems are also planned, and other detections are possible and expected. 

\section*{Acknowledgments}

IXPE is a joint US and Italian mission. The US contribution is supported by the National Aeronautics and Space Administration (NASA) and led and managed by its Marshall Space Flight Center (MSFC), with industry partner Ball Aerospace (contract NNM15AA18C).  The Italian contribution is supported by the Italian Space Agency (Agenzia Spaziale Italiana, ASI) through contract ASI-OHBI-2022-13-I.0, agreements ASI-INAF-2022-19-HH.0 and ASI-INFN-2017.13-H0, and its Space Science Data Center (SSDC) with agreements ASI-INAF-2022-14-HH.0 and ASI-INFN 2021-43-HH.0, and by the Istituto Nazionale di Astrofisica (INAF) and the Istituto Nazionale di Fisica Nucleare (INFN) in Italy.  
This research used data products provided by the IXPE Team (MSFC, SSDC, INAF, and INFN) and distributed with additional software tools by the High-Energy Astrophysics Science Archive Research Center (HEASARC), at NASA Goddard Space Flight Center (GSFC).
V.K. acknowledges support from the Finnish Cultural Foundation.
J.P. thanks the Academy of Finland grant 333112 for support.

\vspace{5mm}
\facilities{IXPE} 

\software {\textsc{ixpeobssim} \citep{Baldini2022}, \textsc{xspec} \citep{Arnaud1996}
}

\bibliography{binaries_paper}{}
\bibliographystyle{aasjournal}

\end{document}